\documentclass[9pt]{article}
\usepackage{geometry}
\usepackage{multicol}
\usepackage{hyperref}
\usepackage[utf8]{inputenc}
\usepackage[T1]{fontenc}
\usepackage{xcolor}
\usepackage{amsmath, amssymb}
\usepackage{graphicx}
\usepackage[round]{natbib}
\usepackage{hyperref}
\usepackage{geometry}
\usepackage{setspace}
\usepackage{authblk}
\usepackage{helvet}
\usepackage[T1]{fontenc} 
\usepackage{enumitem}

\begin{document}
\begin{center}
  \begin{minipage}{0.95\textwidth}
    \centering
    {\LARGE \bfseries Mouse vs. AI: A Neuroethological Benchmark for Visual Robustness and Neural Alignment \par}
    \vspace{1em}

    {\large \bfseries Marius Schneider$^{1,5,*}$, Joe Canzano$^{2,*}$, Jing Peng$^{3,*}$, Yuchen Hou$^{3,*}$, \\
    Spencer LaVere Smith$^{2,5,\dagger}$, Michael Beyeler$^{3,4,\dagger}$ \par}
    \vspace{1em}

    {\small \itshape
    $^{1}$Institute for Collaborative Biotechnologies, University of California, Santa Barbara, CA, USA \\
    $^{2}$Department of Electrical and Computer Engineering, University of California, Santa Barbara, CA, USA \\  
    $^{3}$Department of Computer Science, University of California, Santa Barbara, CA, USA \\  
    $^{4}$Department of Psychological \& Brain Sciences, University of California, Santa Barbara, CA, USA \\
    $^{5}$Correspondence: marius\_schneider@ucsb.edu; sls@ucsb.edu \par    
    $^{*}$ Substantial contribution \par 
    $^{\dagger}$ Equal contribution \par
    }
  \end{minipage}
\end{center}
%
\vspace{1em}
\textbf{
Visual robustness under real-world conditions remains a critical bottleneck for modern reinforcement learning agents. 
In contrast, biological systems such as mice show remarkable resilience to environmental changes, maintaining stable performance even under degraded visual input with minimal exposure.
Inspired by this gap, we propose the Mouse vs. AI: Robust Foraging Competition, a novel bioinspired visual robustness benchmark to test generalization in reinforcement learning (RL) agents trained to navigate a virtual environment toward a visually cued target.
Participants train agents to perform a visually guided foraging task in a naturalistic 3D Unity environment and are evaluated on their ability to generalize to unseen, ecologically realistic visual perturbations. 
What sets this challenge apart is its biological grounding: real mice performed the same task, and participants receive both behavioral performance data and large-scale neural recordings (19,000+ neurons across visual cortex) for benchmarking. 
The competition features two tracks: (1) Visual Robustness, assessing generalization across held-out visual perturbations; and (2) Neural Alignment, evaluating how well agents’ internal representations predict mouse visual cortical activity via a linear readout.
We provide the full Unity environment, a fog-perturbed training condition for validation, baseline proximal policy optimization (PPO) agents, and a rich multimodal dataset. 
By bridging reinforcement learning, computer vision, and neuroscience through a shared, behaviorally grounded task, this challenge advances the development of robust, generalizable, and biologically inspired AI.
}
\vspace{-1em}
\paragraph{Keywords} Robustness, Reinforcement Learning, Generalization under Distribution Shift, Embodied Agents, Biologically Inspired AI, Visual Processing, Neuroscience, Navigation
\begin{multicols}{2}
\section*{Introduction}
%
\vspace{-1em}
Generalization under distributional shift remains a fundamental challenge in machine learning. Visual systems trained under standard conditions often fail when exposed to realistic perturbations such as fog, dynamic lighting, or environmental clutter \citep{wang_survey_2024}. This brittleness limits the deployment of vision-based autonomous systems in safety-critical scenarios, including self-driving vehicles \citep{lu_deepqtest_2023}, drones \citep{gao_weather_2021}, and service robots operating in real-world environments.

In contrast, biological systems exhibit remarkable visual robustness. Mice, for example, can navigate complex environments and perform visually guided tasks with minimal prior exposure, even when visual input is degraded. This discrepancy cannot be explained by superior sensors as biological eyes lag behind modern cameras in resolution and precision, but instead suggests that the visual system implements computational principles that support invariant and noise-tolerant processing. Understanding these principles could help close the gap between biological and artificial vision.

To develop a deeper understanding of how sensory inputs give rise to neural responses and ultimately to robust perception, neuroscience has increasingly turned to predictive models as a powerful tool for probing the visual system. These models span a spectrum of complexity, from interpretable linear-nonlinear cascades to deep learning–based architectures \citep{lurz_deep_2024}. 
While simpler models are easier to analyze, they often lack the capacity to generalize across stimuli or recording conditions. Deep convolutional and transformer-based models offer much higher predictive power, generalizing to novel neurons and stimuli \citep{wang_foundation_2025}, albeit at the cost of interpretability \citep{lurz_deep_2024}. Nonetheless, their success has enabled new insights into how information is processed along the visual hierarchy.
\begin{figure*}[!t]
\vspace{-5mm}
\includegraphics[width=1\textwidth]{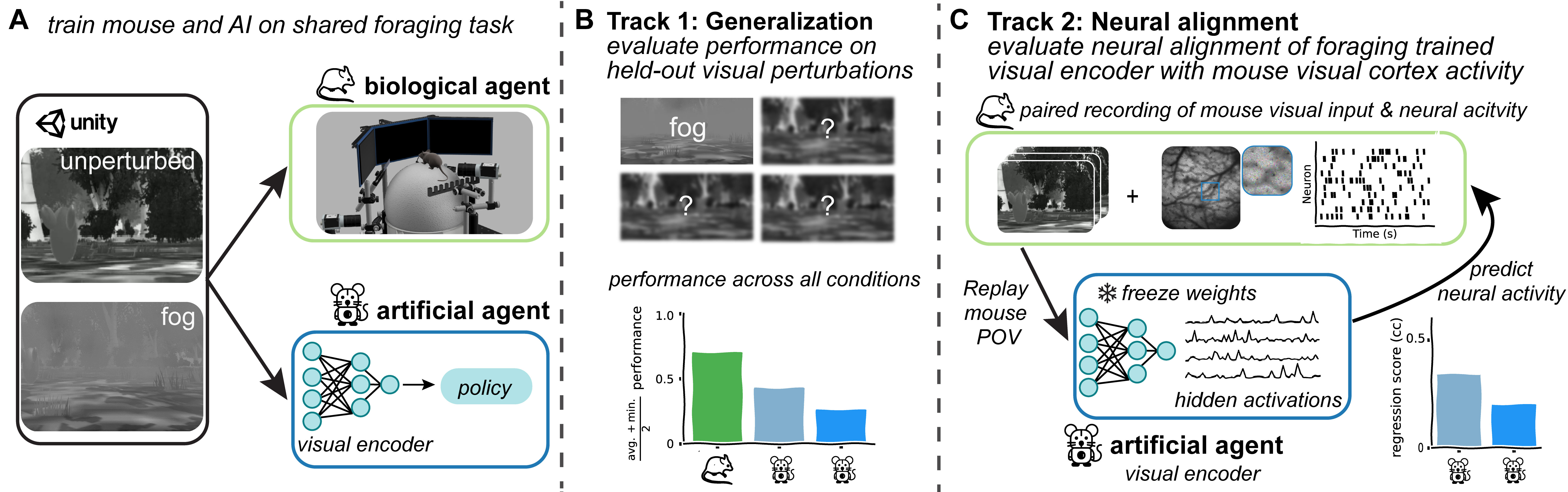}
    \centering
    \vspace*{-3mm}
    \caption[Caption]{{\bf Schematic illustration of the Mouse vs. AI: Robust Foraging competition.}
    \textbf{A.}
     Mice and agents perform the same foraging task, navigating toward a visually cued target in a shared Unity-based environment. 
     Shown are two example frames encountered during behavior: one under normal visual conditions (top) and one under a perturbed condition (bottom). 
    \textbf{B.} 
    Track 1 evaluates an agent’s generalization performance across visual perturbations. 
    Submissions are ranked using two metrics: (1) Average Success Rate (ASR) across all perturbed conditions (including one seen during training), and (2) Minimum Success Rate (MSR) across conditions to assess worst-case robustness. This scoring emphasizes both general competence and failure-mode sensitivity.
    \textbf{C.} 
    Track 2 evaluates spontaneous alignment between the visual encoders of the agent and mouse V1 and HVA brain activity. 
    After training on the foraging task, visual input from mouse behavior is replayed through the agents visual encoder (with frozen weights), and internal activations are extracted. 
    A linear regression is trained to predict recorded neural responses from hidden activations of the visual encoder. 
    Alignment is quantified by the mean Pearson correlation between predicted and observed activity across held-out neurons (regression score). 
    Agent weights remain fixed throughout; only the readout is trained using neural data.
    }
    \label{fig1}
\end{figure*}

Recent architectural innovations have advanced both prediction accuracy and biological interpretability. For instance, Klindt et al. introduced a ``what-where'' separation to disentangle image features from receptive field locations \citep{klindt_neural_2018}, while Ecker et al. incorporated rotation equivariance to reflect cortical symmetries \citep{ecker_rotation-equivariant_2018}. Sinz et al. emphasized the importance of temporal dynamics in modeling neural responses to natural scenes \citep{sinz_stimulus_2018}.

More recently, models have expanded in scale and scope. Wang et al. developed a foundation model trained across multiple mice and modalities, enabling generalization to unseen animals and stimuli \citep{wang_foundation_2025}. Other approaches, such as those \cite{xu_multimodal_2023} and \cite{antoniades_neuroformer_2023}, focused on mice in dynamic, behaviorally relevant tasks, capturing task engagement, locomotion, and sensorimotor contingencies in their predictions.

To systematically compare these models and drive community progress, large-scale benchmarks have emerged. Brain-Score \citep{schrimpf2018brain} evaluates how well internal representations of task-trained models align with neural activity across species and regions. Sensorium \citep{willeke_sensorium_2022, turishcheva_retrospective_2024}, in contrast, provides raw population recordings from mouse V1 and invites participants to train encoding models directly on these data.

These benchmarks have played a central role in advancing model–brain alignment. However, both are limited to passive viewing conditions. It is well-established that neural activity in visual cortex is strongly modulated by movement, task engagement, and behavioral context \citep{parker_movement-related_2020, saleem_interactions_2023, niell_modulation_2010, minderer_spatial_2019}. As a result, benchmarks based on passive stimuli may capture only a subset of the computations that emerge during naturalistic, goal-directed behavior. Furthermore, training models directly on neural data requires large, high-entropy datasets to avoid overfitting, an ongoing challenge in sensory neuroscience.

To address these limitations, we introduce the \textbf{Mouse vs. AI: Robust Foraging} competition: a benchmark that jointly evaluates \textbf{visual robustness} and \textbf{neural alignment} in artificial agents. Agents are trained to perform a visually guided foraging task in a 3D Unity environment and are evaluated on their ability to generalize to a set of visual perturbations. Crucially, head-fixed mice perform the same task in the same environment via a floating-ball setup, enabling comparisons grounded in shared visual input and task structure.

Large-scale two-photon calcium imaging and behavioral data from these mice are used to evaluate representational alignment. Importantly, submitted models are trained only to solve the task; alignment with biological data is assessed post hoc via a linear readout from fixed model activations. This design allows us to test whether brain-like representations can emerge spontaneously through behavior-driven learning without using any neural data during training.

Unlike previous efforts that isolate robustness (e.g., ImageNet-C \citep{hendrycks2019benchmarking} and the Adversarial Vision Challenge \citep{brendel2020adversarial}) or model-brain alignment (e.g., Sensorium and Brainscore \citep{schrimpf2018brain,willeke_sensorium_2022,turishcheva_retrospective_2024}), our benchmark combines both in a unified, biologically-grounded setting. Ultimately, the benchmark enables researchers to test whether robustness and brain-like representations co-emerge through behavior-driven learning, or whether neural alignment demands dedicated architectural or training constraints.
\begin{figure*}[!t]
\vspace{-5mm}
\includegraphics[width=1\textwidth]{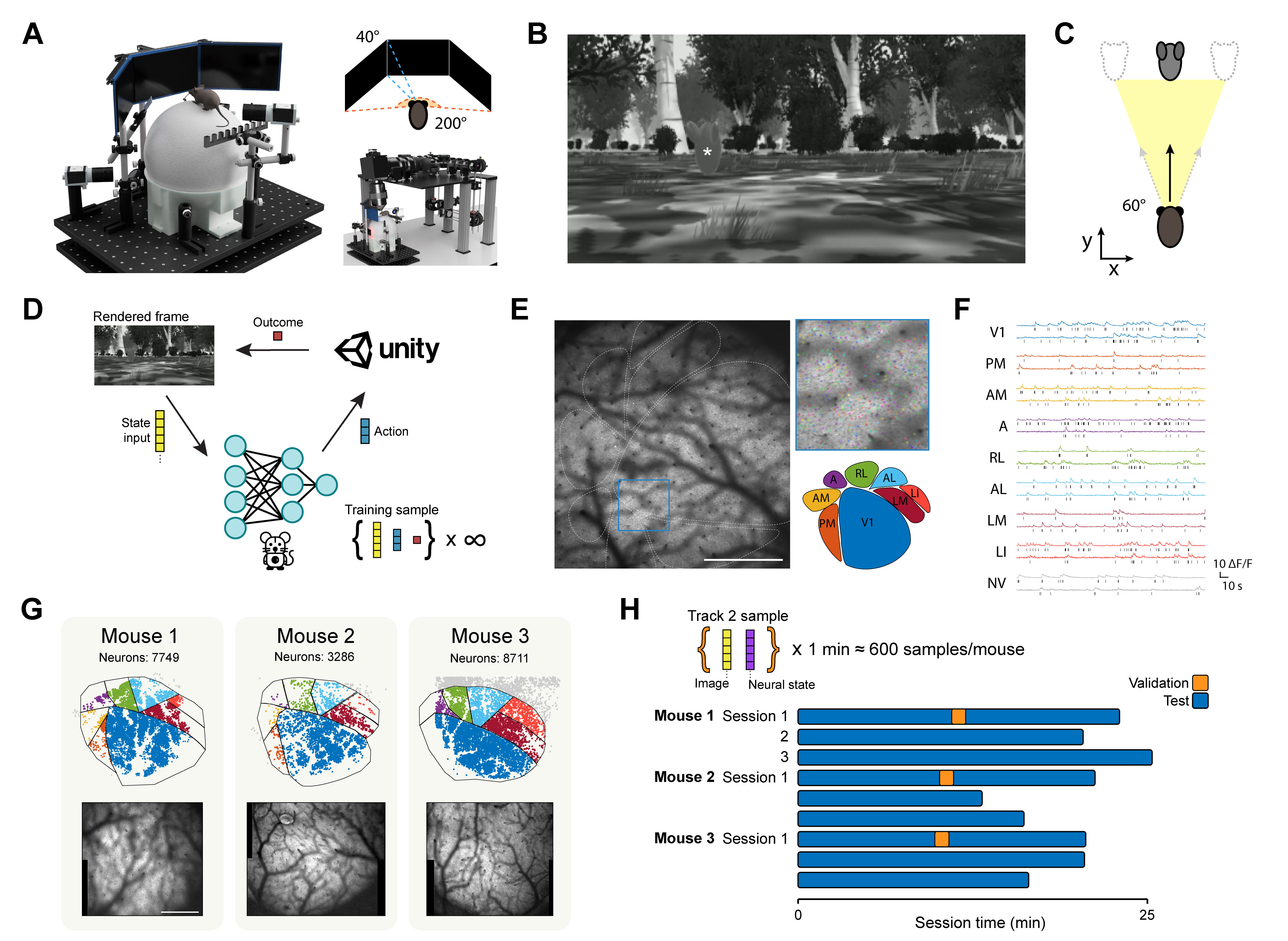}
    \centering
    \vspace*{-3mm}
    \caption[Caption]{{\bf Task, virtual environment, and dataset details.}
    \textbf{A.} Mouse VR setup. The mouse is mounted on an omnidirectional treadmill and an array of screens (top right) provides visual feedback. Bottom right, to collect track two data, expert mice performed the virtual task concurrent with mesoscale two-photon imaging with the Diesel2P system. 
    \textbf{B.} 
    Appearance of virtual environment. Target object labeled with white marker. 
    \textbf{C.}    
    Each trial, the target object spawns at a set distance with a continuous random left/right offset. It always spawns in the field of view (marked in yellow). The reward condition is triggered when the player contacts the target. 
    \textbf{D.} Track one training procedure and training sample composition.
    \textbf{E.} Example imaging field of view (FOV) with segmented cell masks. Scale bar, 1 mm.
    \textbf{F.} Example calcium traces colored by visual area and inferred spike trains (black) from individual neurons.
    \textbf{G.} Spatial maps of cell locations across cortical visual areas for each mouse. Bottom, full imaging FOV projections. Scale bar, 1 mm.  
    \textbf{H.} Track two data overview. Top, each sample consists of one paired image and neural state. Bottom, validation and test datasets from all mice and sessions.
    }
    \label{fig1}
\end{figure*}

\section*{Competition Overview}
%
The goal of the Mouse vs. AI: Robust Foraging competition is two-fold. 
First, it assesses the ability of artificial agents to generalize across naturalistic visual perturbations in a behaviorally relevant setting. 
Second, it evaluates whether internal representations developed through task training spontaneously align with neural activity recorded from the visual cortex of mice performing the same task.
Both objectives are addressed within a unified benchmark that pairs a shared visual foraging task with a multimodal biological dataset. 
This design enables systematic study of the principles that support robust, biologically inspired vision.

The competition is organized into two complementary tracks built around a shared visually guided navigation task in a naturalistic 3D Unity environment.
Participants train RL agents to navigate to a target object based solely on egocentric visual input. During training, they have access to the environment under normal conditions and a single example perturbation (fog). After training, submitted models are evaluated on hidden test data in two tracks (\url{https://robustforaging.github.io/leaderboard/}) :
\begin{itemize}[topsep=0pt, itemsep=0pt, parsep=0pt]
    \item \textbf{Track 1: Robustness.} Track 1 evaluates each agent's success rate on held-out visual perturbations. The goal is to develop RL agents that can navigate robustly under distributional shifts.
    \item \textbf{Track 2: Neural Alignment.} Track 2 evaluates neural alignment by feeding mouse visual input into each trained model. A linear readout is trained to predict recorded neural activity, and performance is measured as the correlation between the predicted and observed responses. The goal is to explore alignment between artificial models and cortical representations.
\end{itemize}

Agents must be trained solely to solve the navigation task; no neural data may be used to modify their representations. Alignment is assessed post hoc to test whether brain-like representations can emerge spontaneously from behavior-driven learning.
To ensure neural alignment is assessed on competent agents, models considered for Track~2 must also achieve adequate task performance (avg. success rate $>0.5$ at the maximum target distance) in the validation environment.

This structure mirrors prior model-brain benchmarks such as Sensorium and Brainscore \citep{schrimpf2018brain,willeke_sensorium_2022, turishcheva_retrospective_2024}, but differs in two key ways: (i) visual input is embedded in an active sensorimotor loop, rather than passive viewing, and (ii) the same environment is used for both artificial agents and biological subjects, enabling direct comparison of behavior and neural responses under matched conditions.

A starter kit provides the Unity environment, baseline agents, data loaders, and evaluation scripts to streamline participation. All submissions are containerized and evaluated on held-out test data by the organizers.

\subsection*{Visual Foraging Task}
%

The benchmark centers on a visually guided navigation task implemented in the Unity game engine . At each timestep, the agent receives an egocentric grayscale frame as input and must navigate toward a cued target object using vision alone. 
Mice performed the same task in virtual reality, ensuring a direct comparison between artificial and biological agents (Figure~2A, see Methods for details).

To increase visual complexity and ecological relevance, the environment was populated with naturalistic features that approximate real-world visual statistics and introduce trial-to-trial variability (Figure 2B). Key features include:

\begin{itemize}[topsep=0pt, itemsep=0pt, parsep=0pt]
\item  Trees with textured trunks and leafy canopies
\item  Shrubs and grass of varying sizes and depth
\item  Ground textures that yield naturalistic optic flow patterns during movement
\item  Dynamic lighting with cast and ambient shadows
\item  Wind-driven motion of foliage and shadows
\end{itemize}

In each trial, a gray target object spawns at a distance with a random horizontal offset (uniformly distributed within the range $\pm$30°). The agent must reach the target within five seconds to be counted as a \textit{success}; otherwise, the trial ends in \textit{failure}. Target distance increases with performance until a maximum is reached (approx. 0.8 m), then remains fixed for the session. The target had similar luminance to its surroundings, reducing its salience and making it nontrivial to segment from the environment. Mice typically required 13 training sessions to learn the task, with experts routinely achieving  >70\% performance over 200 - 300 trials and approximately 20 minutes of active behavior.

\section*{Data}
The benchmark provides two complementary datasets: Unity builds with mouse behavioral baselines (Track 1), and paired video-neural recordings from mice performing the same task (Track 2).
\subsection*{Track 1: Unity Builds}
Participants train their models via online reinforcement learning in a Unity-based simulator. No pre-collected training dataset is provided; instead, agents must collect their own experience 
by interacting with the environment.

At each timestep $t$, the agent receives a grayscale egocentric visual observation $s_{t} \in \mathbb{R}^{86 \times 155}\ $ and outputs a continuous action $a_{t} \in \mathbb{R}^3$ representing forward/backward translation, lateral translation, and rotation. 
The environment executes the action, updates its internal state, and returns the next observation $s_{t+1}$ along with a scalar reward:
\[
r_t = 
\begin{cases}
+20, & \text{if the target is reached}, \\
-1,  & \text{otherwise}.
\end{cases}
\]
Together, these elements define the reinforcement learning transition \(\{s_t, a_t, r_t,s_{t+1}\}\).

\textbf{Training and validation environment (provided).\quad} Participants are provided with three Unity builds for training and local validation:
\begin{itemize}[topsep=0pt, parsep=0pt, itemsep=0pt]
    \item Normal – trials without perturbations
    \item Fog – trials with fog perturbation only
    \item Random – normal and fog trials interleaved
\end{itemize}
Each build comes in two versions:
\begin{itemize}[topsep=0pt, parsep=0pt, itemsep=0pt]
    \item a training version, where target distance increases with success rate until a maximum is reached;
    \item a validation version, where targets start at the maximum distance from the beginning.
\end{itemize}
\textbf{Test environment (hidden).\quad} During public evaluation, each submitted model is tested on 100 trials each from normal, fog, and three additional held-out perturbations, totaling 500 trials per model. 
As a biological baseline, we also provide trial-averaged success rates from $n=3$ expert mice performing the same task across six behavioral sessions (mean $\sim$210 trials per session), averaged separately for each perturbation condition.

\subsection*{Track 2: Neural Activity}
While mice performed the same foraging task in virtual reality, mesoscale two-photon imaging was used to record population activity from layer 2/3 neurons in primary visual cortex (V1) and multiple higher visual areas (HVAs)(Figure 2E–G). See \emph{Methods} for recording details.

The dataset consists of paired visual input and neural recordings:  
\[
\{(S_{1:T}, N_{1:T})\}, \quad S_{1:T}\in\mathbb{R}^{86\times155\times T},\; N_{1:T}\in\mathbb{R}^{n\times T}.
\]
Here, $S_{1:T}$ denotes the sequence of $T$ egocentric video frames, where each frame $S_t\in\mathbb{R}^{86\times155}$ is a single grayscale image, and $N_{1:T}$ denotes the corresponding neural activity from $n$ simultaneously recorded neurons over the same $T$ time steps.

\textbf{Validation set (provided).} Three short samples ($\sim$1~min each) from three mice, covering 19{,}746 neurons in total (Figure~2G--H). This set is intentionally small and provided only for model selection and candidate comparison. 
Using neural data to train or modify agent representations is not permitted and would result in overfitting.

\textbf{Test set (hidden).} Nine additional sessions ($\sim$178~mins total), covering more than 50{,}000 neurons across up to eight cortical visual areas, are held back for the official evaluation of submitted models. These data are not accessible to participants.  

For evaluation, we replay the recorded visual sequences $S_{1:T}$ through each submitted model and extract hidden-layer activations $M_{1:T}$. A linear readout is then fit to predict the recorded neural activity $N_{1:T}$ from these activations. 
Here, ``\emph{post hoc} linear readout'' means that agent parameters are frozen; only a linear mapping from model activations to neural responses is fit for evaluation.
The precise procedure and performance metric are described in the \emph{Metrics} section.

\section*{Metrics}

Each track uses a distinct, well-defined evaluation metric aligned with its scientific objective. All final scores will be computed by the organizers using centralized infrastructure on held-out test data. Participants will submit trained models and inference code via a containerized environment.

\subsection*{Track 1: Robustness Evaluation}
Robustness is evaluated based on agent performance across five visual conditions: the normal and fog environments (provided for training) and three held-out perturbations (hidden). A trial is counted as successful if the agent reaches the target within the 5-second time frame.  

Each agent is evaluated over 500 trials (with randomized seeds) in each condition. Two metrics are reported:  
\begin{itemize}[topsep=0pt, parsep=0pt, itemsep=0pt]
    \item \textbf{Average Success Rate (ASR):} mean success rate across all perturbation conditions.  
    \item \textbf{Minimum Success Rate (MSR):} lowest success rate across all conditions, capturing worst-case generalization.  
\end{itemize}  

Final leaderboard ranking is based on a weighted combination:  
 \[
\mathrm{Score}_{1} = 0.5 \cdot \mathrm{ASR} + 0.5 \cdot \mathrm{MSR}
\]

\subsection*{Track 2: Neural Alignment Evaluation}

Alignment is quantified by fitting a linear mapping from fixed model activations to recorded neural responses; the score is the mean Pearson correlation across neurons (reported for the best layer).
%
For each submitted model, we extract activations from every layer $l \in \{1, \dots, L\}$. Let  
\[
M_t^{(l)} \in \mathbb{R}^{n \times d}
\]  
denote the activations Matrix of layer $l$ (with $n$ samples and $d$ features). 
To control for differences in the dimensionality of the representation, each activation $M_t^{(l)}$ is projected onto a lower-dimensional subspace using principal component analysis (PCA):
\[
\ \tilde{M}^{(l)} = M^{(l)} V_k, \quad V_k \in \mathbb{R}^{d \times k},\; V_k^\top V_k = I_k,
\]
where $V_k$ contains the top $k$ principal components (eigenvectors of the covariance matrix of $M$).
The resulting matrix $\tilde{M}^{(l)}  \in \mathbb{R}^{n \times k}$ is the reduced representation of M.
A linear regression is then fit from $\tilde{M}_t^{(l)}$ to the recorded neural responses $N_t$:  
\[
\hat{N}_t^{(l)} = W^{(l)} \tilde{M}_t^{(l)} + b^{(l)},
\]  
where $W^{(l)}$ and $b^{(l)}$ are fitted parameters.  \\
For each layer $l$, performance is quantified as the mean Pearson correlation across neurons,  
\[
\rho^{(l)} = \frac{1}{n} \sum_{i=1}^n \text{corr}\!\left(\hat{N}_{t,i}^{(l)}, N_{t,i}\right),
\]  
where $N_{t,i}$ is the response of neuron $i$.  
The \textbf{final alignment score} of a model is defined as the highest average correlation across all layers,
\[
\text{Score}_{2} = \max_{l \in \{1,\dots,L\}} \rho^{(l)}.
\]

\section*{Baselines and Starter Kit}
\label{sec:baseline}
To lower the barrier to entry, we provide a complete starter kit on 
\href{https://github.com/robustforaging}{github.com/robustforaging} 
under a permissive open-source license. It includes:
\begin{itemize}[topsep=0pt, itemsep=0pt, parsep=0pt]
  \item \textbf{Unity Environment:} Clean and fog-perturbed environments with a Python API 
        built on the open-source \href{https://unity-technologies.github.io/ml-agents/ML-Agents-Overview/}{ML-Agents} framework.
  \item \textbf{Baseline Agents:} PPO-trained agents implemented with ML-Agents, combining a visual encoder 
        with a two-layer MLP policy head that maps features to continuous actions. 
        Provided encoder backbones include a single fully connected (FC) layer, a 3-layer CNN, 
        a ResNet-18 \citep{he_deep_2015}, and a neuro-inspired CNN previously shown to predict mouse neural activity \citep{xu_multimodal_2023}. 
        All agents are trained from scratch on the normal and fog environments using identical PPO hyperparameters.
  \item \textbf{Evaluation Utilities:} Scripts to load and preprocess trial-level behavior 
        and two-photon neural recordings, extract model activations, and fit the linear decoder for Track~2.
  \item \textbf{Reproducibility Toolkit:} Conda environment, Python scripts, and Jupyter notebooks 
        demonstrating end-to-end setup, training, and evaluation.
\end{itemize}

Baseline agent performance (average success rate on clean and fog) is provided as a reference point 
for participants to reproduce and improve upon. All code, data, and environments are version-controlled 
to guarantee that results are fully reproducible.

\section*{Discussion}
The Mouse vs. AI: Robust Foraging competition was designed with two primary goals: to evaluate the visual robustness of artificial agents under distributional shift, and to assess whether their internal representations align with neural activity recorded from mice performing the same task. Unlike traditional benchmarks that focus on either task generalization or neural predictivity, our competition bridges both challenges within a unified, biologically grounded framework. By training and evaluating agents in the same naturalistic 3D environment used for mouse behavior and neural recordings, the benchmark enables principled comparisons across artificial and biological systems, linking perception, action, and neural representation.

Track 1 of the competition focuses on evaluating the visual robustness of artificial agents trained to perform a biologically inspired foraging task. Agents are optimized under standard conditions and one perturbation type (fog), and then evaluated on their ability to generalize to several held-out visual perturbations that introduce a naturalistic distributional shift. Crucially, head-fixed mice were trained to perform the same task in the same Unity-based virtual environment, allowing for direct comparisons between species under matched perceptual and behavioral conditions. 
Unlike prior robustness benchmarks such as ImageNet-C \citep{hendrycks2019benchmarking} or the Adversarial Vision Challenge \citep{brendel2020adversarial}, which evaluate static image classification under synthetic corruptions, our benchmark embeds these perturbations in an active sensorimotor loop. Initial results reveal a robustness gap: mice maintain high success rates under challenging conditions, while baseline agents often fail, underscoring the need for better inductive biases and training strategies in artificial systems.

Track 2 evaluates the extent to which artificial agents develop internal representations that align with neural activity recorded from mice performing the same task. Participants submit agents trained solely to solve the visually guided foraging task, and alignment is evaluated post hoc via linear regression from fixed model activations to two-photon calcium recordings. This approach avoids the risk of overfitting limited neural datasets, which is an inherent challenge in traditional encoding models trained directly on neural activity, and instead mirrors the strategy used in benchmarks like Brain-Score \citep{schrimpf2018brain}, where models are optimized for a separate task and evaluated afterward for neural alignment.
However, while Brain-Score primarily benchmarks models trained on high-level object recognition—an ecologically relevant function for primate and human vision—such tasks are less aligned with the rodent visual system, which is thought to be more specialized for motion processing, spatial navigation, and ethologically relevant sensorimotor behaviors \citep{saleem2013,saleem2018,parker_movement-related_2020,flossmann2021,SKYBERG2024}. 
By contrast, the Mouse vs. AI benchmark evaluates models trained on a biologically inspired foraging task in a naturalistic 3D environment in which vision is directly used to guide movement and decision-making. This setup offers a more ecologically valid framework for testing neural alignment in mouse visual cortex, and provides a principled testbed for investigating whether brain-like representations emerge naturally from solving tasks that reflect the behavioral demands and sensory specializations of the species.
%

Our competition aims to provide insights into how architectural modifications and training strategies influence models' path-finding performance and their neural alignment during spatial navigation. Prior studies have shown that models with more brain-like features, such as embedded internal states and gain factors (e.g., Sensorium 2022 behavior track top 1), and increased architectural complexity (e.g., Sensorium 2023) tended to achieve stronger neural prediction and greater robustness to out-of-distribution stimuli. Similarly, models trained on richer input diets \citep{conwell2024}, with self-supervised learning \citep{bakhtiari2021,nayebi2023}, or using tasks relevant to navigation \citep{conwell2021} demonstrated superior neural alignment.
However, it remains unclear whether these features continue to play a critical role in a more dynamic and ethologically relevant context, such as the one in our competition, or whether additional factors that are often overlooked in passive viewing conditions will emerge as equally or more important. By creating this competition, we seek to directly test these hypotheses and uncover which architectural and training principles truly matter for robust navigation models.

The benchmark provides a rich multimodal dataset linking visual input, behavior, and large-scale neural activity across multiple visual areas, including V1 and higher-order regions. 
Whereas most predictive modeling efforts rely on data collected during passive viewing of static or dynamic stimuli (e.g., Sensorium \citep{willeke_sensorium_2022,turishcheva_retrospective_2024}), our recordings are acquired during active task performance. 
This enables analysis of visual processing in behaviorally relevant contexts where vision is used to guide navigation and goal-directed actions, rather than simply respond to presented stimuli.
As a result, the dataset captures not only visual responses, but also task-related modulations that shape neural activity in real-world contexts.
By embedding neural activity within a goal-directed task, the benchmark provides a biologically grounded testbed for evaluating representational alignment under realistic, ethologically meaningful conditions, thereby pushing beyond traditional benchmarks that isolate stimulus-driven responses from action or behavioral context.

A key strength of the Unity-based environment is its ability to manipulate visual input with experimental precision, which introduces perturbations such as fog or lighting changes while keeping task demands constant. This enables systematic evaluation of model robustness under different types of distributional shift and facilitates hypothesis-driven testing of architectural inductive biases, training regimes, or data augmentation strategies. It also supports causal analyses in artificial agents (e.g., by lesioning network units or altering reward structures) to dissect how specific components contribute to behavior and representation. Finally, the shared task structure between mice and agents supports behavioral fingerprinting: comparing how different systems respond to identical sensory and decision-making demands to reveal convergent or divergent strategies.

These properties enable a wide range of future extensions. Visual perturbations could be expanded to include additional forms of sensory degradation or distributional shift. Task complexity could be increased with multi-step objectives, distractor targets, or shifting reward contingencies that probe flexibility and planning. Additional behavioral motifs (such as predator-like stimuli or dynamic prey) could simulate ecologically relevant challenges and allow researchers to probe how visual information is processed during rapid decision-making in realistic scenarios. The Unity-based environment offers a unique combination of biological realism and experimental control, enabling the design of naturalistic behavioral challenges with precisely manipulated visual inputs, while preserving reproducibility and interpretability.

By grounding vision in action and jointly evaluating robustness and neural alignment within a unified behavioral framework, the Mouse vs. AI benchmark represents an important step toward biologically inspired machine learning. We hope it will serve as a foundation for continued research at the intersection of neuroscience, reinforcement learning, and computer vision.

\section*{Acknowledgments}
%
This work was supported by the National Institute of Neurological Disorders and Stroke of the National Institutes of Health under Award Number R01-NS121919. The content is solely the responsibility of the authors and does not necessarily represent the official views of the National Institutes of Health.
\section*{Author Contributions}
%
\textbf{MS:} Conceptualization, Formal Analysis,
Investigation, Methodology, Project Administration, Software, Data Curation, Validation, Visualization, Writing - original draft, Writing - Review and Editing
\textbf{JC:} Conceptualization, Formal Analysis, Investigation, Methodology, Animal Experiments, Software, Data Curation, Validation, Visualization, Writing - original draft, Writing - Review and Editing
\textbf{JP:} Investigation, Software, Validation, Methodology, Project Administration, Visualization, Writing - Review and Editing
\textbf{YH:} Conceptualization, Formal Analysis,
Investigation, Methodology, Project Administration, Software, Validation, Writing - original draft, Writing - Review and Editing
\textbf{SLS:} Conceptualization, Methodology,
Funding Acquisition, Supervision, Writing - Review and Editing
\textbf{MB:} Conceptualization, Methodology,
Funding Acquisition, Supervision, Writing - Review and Editing

\section*{Methods}
\subsection*{Experiments}

\paragraph{Animals}
All animal procedures and experiments were approved by the Institutional Animal Care and Use Committee of the University of California Santa Barbara and performed in accordance with US Department of Health and Human Services regulations. This study includes recordings from three adult mice including both sexes, aged 18 to 50 weeks, utilizing the TIGRE locus for pan-neural expression of the fluorescent calcium indicator GCaMP6s \citep{chen_ultra-sensitive_2013} in cortex. The exact genotype was a triple cross between TITL-GCaMP6s (Ai94, \citep{madisen_transgenic_2015}), Emx1-Cre (Jackson Labs \#005628), and ROSA:LNL:tTA (Jackson Labs \#011008). Mice were housed in a 12h/12h reversed light cycle throughout the experiments.

Cranial windows were implanted over the right posterior cortex in mice prior to imaging. Mice were anesthetized with 1 - 2 \verb|%| isoflurane gas and maintained at 37$^\circ$C body temperature throughout the procedure with a closed loop heating pad. Systemic carprofen (10 mg/kg) and local lidocaine (10 mg/kg) were provided for analgesia, and ophthalmic ointment was applied to both eyes to prevent desiccation. The portion of skin and periosteum dorsal to the skull were resected, and a 4 mm diameter craniotomy was performed over visual cortex guided by bone landmarks. The craniotomy was closed with a plug consisting of two glass coverslips of \verb|#|1 thickness adjoined with optical UV resin, and adhered in place with cyanoacrylate glue (Oasis Medical). A stainless steel headplate with 7 mm aperture was then adhered around the preparation with dental cement (Parkell Metabond). Animals received 2-4 days of postoperative oral carprofen (20 mg/kg) and were allowed a full week of recovery before water restriction or behavior training.

\paragraph{Intrinsic signals optical imaging (ISOI)}
After recovery, ISOI \citep{kalatsky_new_2003} was performed to measure cortical retinotopic maps used to delineate visual area boundaries as described previously \citep{yu_selective_2022,smith_stream-dependent_2017}. Briefly, animals were lightly anesthetized with isoflurane (0.25 - 0.8 \verb|%|) and acepromazine (1 - 3 \verb|%| mg/kg) and maintained at 37 $^\circ$C with a closed loop heating pad. A tandem lens macroscope with a 4.7 mm\textsuperscript{2} field of view, CCD camera (Teledyne DALSA 1M30), and broadband halogen light source imaged reflectance signals in either green (Ex: 550 ± 50 nm, Em: 560 ± 5 nm, Edmund Optics) or red (Ex: 700 ± 38 nm, Em: 700 ± 5 nm, Edmund Optics). Vasculature reference images were collected in green at the cortical surface, then ISOI videos were recorded in red at 600 $\mu$m depth at 30 Hz. During ISOI recordings, visual stimuli were provided with a 60 $\times$ 34 cm\textsuperscript{2} flat monitor from a distance of 20 cm, covering 110° $\times$ 75° in azimuth, elevation the visual field for the left eye. Periodic whole field drifting bar stimuli were used to collect azimuth and elevation phase maps separately, and area boundaries were drawn by hand using gradient reversal patterns and other landmarks \citep{marshel_functional_2011,smith_stream-dependent_2017}. The vasculature references were then used to register retinotopic coordinates with subsequent two-photon imaging.

\paragraph{VR apparatus}

Mice were engaged in head-fixed visual navigation using a custom VR apparatus. Broadly, this included a screen array, an omnidirectional treadmill constructed from a 9 inch hollow styrofoam ball suspended on air \citep{dombeck_imaging_2007,dombeck_functional_2010} and 3D printed base, two optical mouse sensors to track the treadmill movement, and a lickport for liquid reward delivery. All rotational degrees of freedom on the treadmill were used in game: pitch rotations mapped to forward/backward translation, roll mapped to left/right translation, and yaw mapped to changes in heading direction. Microcontrollers (Arduino) were used for hardware control and i/o, serial communications with Unity, triggering for synchrony, and capacitive lick sensing. Mice were suspended atop the ball apex with sturdy optomechanics (Thorlabs) interfacing with the implanted headplate, and oriented for their heading direction to be centered on the screen array. Liquid reward delivery was gravity driven and controlled with an inline valve; pulses were calibrated for a single reward volume of 5 $\mu$L. 

Visual feedback was presented with an array of three 7" LCD screens laterally arranged around the animal midline with 60$^\circ$ between screens to symmetrically cover 200$^\circ$ azimuth and 40$^\circ$ elevation in visual angle from the center point between both eyes. The distance from the left eye to the front screen was 11.5 cm and the rendered frames from the game were presented at full 600 x 1024 px resolution, thus 16 px/$^\circ$ in spatial resolution. All frames were rendered in grayscale at full contrast with a frame update frequency of 45 Hz.

Concurrent with behavior, the left eye and pose of the animals were imaged with high speed infrared CMOS cameras (Basler) controlled by custom LabVIEW code. Animals were illuminated with an IR diode light source. One camera recorded eye movements at 30 Hz through a periscope and dichroic shortpass mirror (700nm cutoff, Edmund) situated between the animal and the screen. The other was fitted with a IR longpass filter (750nm cutoff, Edmund) to reject visible screen emission and imaged the whole side profile of the animal reflecting off the left stimulus screen at 72 Hz. Eye and pose camera imaging are synchronized with other signals via TTL (see two-photon imaging section).


\paragraph{Virtual task}

The VR task was implemented in the Unity game engine with custom C\verb|#| scripts. Much of this is described in the main text, see Virtual Foraging Task. Some additional details are as follows. The screen array feedback is rendered from three in game cameras attached to the player object with the same spatial orientation as the screens, each set to perspective projection with the field of view angle matching the mouse's view of the screen. Ball movement measurements from the optical mouse sensors are linearly mapped to 2D translation and 1D rotation coplanar with the ground, with gain parameters set empirically for animal performance. Once set these gains corresponded to full target distances of 0.77 m in expert sessions. 

At full distance the target object occupied 5.5$^\circ$ in azimuth and 10$^\circ$ in elevation. The target left/right offset scaled linearly with distance such that at the goal distance of 0.77 m the offset range is $\pm$ 30$^\circ$ from the center in the visual field. The target object follows the tripod design featured in several other animal studies of visual object recognition \citep{zoccolan_rodent_2009,djurdjevic_accuracy_2018,froudarakis_object_2020}, with its luminance was set to middle gray. While we can't describe the perturbations in detail without spoiling the competition, all were implemented either as particle fields in Unity or as custom HLSL shaders. 


\paragraph{Animal behavior training}

Animals were motivated to perform this task via water restriction to 85\verb|%| baseline weight starting one week prior to behavior training. In successful trials, rewards were 5-15 $\mu$l of water per trial, delivered immediately after the target object is reached, and paired with an audible relay tick which served as a reward cue. There was no punishment for failure trials other than a 250 ms negative 10 KHz tone cue, delivered when the trial timer expired. 

In animal training, each proceeded through a sequence of phases in the task before reaching expert: 
\begin{itemize}
    \item Phase 0: No target object. Animals freely navigate in the environment. Rewards are randomly delivered at 10 $\pm$ 5 second intervals to build association between liquid reward and reward cue, and to acclimate to the process.   
    \item Phase 1: Target spawned straight ahead and within-session distance increases with running performance. The distance increment in turn is increased across sessions. Animals must reach a threshold target distance of 0.77 m and achieve 70\verb|%| session performance to proceed to next phase. 
    \item Phase 2: Target left/right offsets engaged, and otherwise the same sequence as phase 1 is applied. Mice are considered experts on the task once they reach a threshold target distance of 0.77 m and achieve 70\verb|%| session performance.
\end{itemize}

In our experience this process takes about 13 training sessions to complete start to finish (12.85 $\pm$ 0.7 [mean $\pm$ SEM], n=14 mice).

Once mice reach expert performance in phase 2 they proceed to imaging. Between imaging sessions, mice are maintained at expert performance with several maintenance sessions per week on phase 2. In concurrent imaging and behavior sessions, only those reaching the expert criteria for performance are retained. For each mouse, at least three imaging sessions with expert behavior performance were collected.

\paragraph{\textit{In vivo} two photon imaging}

Two-photon imaging during behavior was performed with the Diesel2P (D2P) fluorescence mesoscope \citep{yu_diesel2p_2020} utilizing both independent temporally multiplexed 8 KHz RGG scan engines in the 5 mm objective FOV. Imaging of GCaMP6s expressing layer 2/3 neurons was performed at a depth of 150-250 $\mu m$. The objective focal plane was registered to the craniotomy plane by rotating the objective on one axis, and rotating the VR setup on the coplanar axis with the table. Excitation was provided by a 2W 80 MHz Ti:Sapphire laser (Spectra-Physics MaiTai) at 920nm, with illumination power between 60-80 mW at the sample. In each imaging path, scanning was performed in resonant mode sampling 1024 px on the fast axis over a 1.5 mm field for a spatial resolution of 1.4 $\mu m$/px. During dual path imaging, each scanned a rectangular field of 1.5 mm $\times$ 3-4 mm and the two were placed side by side for a combined field of 3 $\times$ 3-4 mm scanned at 6-8 Hz. PMT signals were demultiplexed using conservative < 3 ns windows for emission detection following laser pulses to limit crosstalk. Visible emission from the VR rig screens was blocked with a custom light block cone which forms a flexible opaque seal between the implanted headplate and objective.  

To synchronize all data recorded during concurrent imaging and behavior, an external timing DAQ (National Instruments) recorded analog or digital signals from all independent hardware on one shared 1 KHz clock. These included: one slow scanner control signal for each D2P imaging path, new trial timestamp triggers from Unity, and eye and pose camera exposure TTL triggers. Full resolution video records of the VR stimulus were captured during behavior with custom python code (DxCam, github.com/ra1nty/DXcam) at 60 Hz from the two screens contralateral to the imaged hemisphere, and synchronized with other signals using the shared system clock with the VR behavior data. VR behavior and imaging sessions were ended once the animals become satiated and running performance declined, usually after 150-300 trials or 15-25 minutes.

\paragraph{Imaging data preprocessing}

Calcium videos from Scanimage for each imaging path were motion corrected and segmented into cellular fluorescence traces with Suite2P \citep{pachitariu_suite2p_2017}. Segmentation was aided by a custom cell classifier in Suite2p trained on our own mesoscale data, and performance was verified by eye on all recordings. The \verb|threshold_scaling| parameter was conservatively set to 1.5 - 2 to reject low SNR ROIs. Raw calcium traces were baseline and neuropil subtracted to generate $\Delta F$ traces via 

\[\Delta F_{t}=F_t-0.7F_{neu,t}-F_{base,t}\]

Where $F$ and $F_{neu}$ are the averaged fluorescence within the cell and surrounding neuropil masks, and $F_{base}$ is a moving baseline calculated by first gauss filtering $F-F_{neu}$ with a 30 second window, then sequentially filtering again with moving minimum and maximum filters in a 120 second window, similar to the stock $\Delta F/F$ calculation in Suite2P. Spike inference was then performed on the $\Delta F$ traces with a bayesian Markov-chain Monte Carlo (MCMC) method which estimates the posterior distribution for spike probability \citep{pnevmatikakis_bayesian_2013}. As a quality control step, for each cell, 400 samples of discrete spike trains were generated from the posterior and used to estimate inference reproducibility. After binning each in non-overlapping 100 ms windows, the mean Pearson correlation was computed between all samples (including tens of minutes per recording), and only cells passing a > 0.4 threshold were retained in all following analysis. The number of neurons which passed this criterion was defined as the total number of neurons. For each included cell, one spike train sample was chosen at random for all subsequent analysis. Both imaging paths were processed with the above steps independently, then registered to ISOI coordinates via vascular landmarks to assign cells to visual areas.

\paragraph{Screen and neural data preprocessing}

Finally, the neural and screen data from the VR experiments were preprocessed for ease of use, formatted as samples $\{S_t,N_t\}$ pairing each stimulus frame $S_t\in \mathbb{R}^{x,y}$ at time $t$ with a neural state $N_t\in \mathbb{N}^n$. To do this, spike trains per cell were binned in non-overlapping 100 ms windows and associated with the stimulus frame on the center screen presented 60 ms prior to account for neural latency. Stimulus frames were then average downsampled to 150 $\times$ 256 px pixels in size representing 0.25 $^\circ$/px and roughly 4x mouse visual acuity. This resolution surplus is provided in case it is useful; competitors may wish to simply average downsample again to match the RL agent screen resolution.\footnote{We find the combination of 100 ms bin width and spatial resolution anywhere between 4x - 1x of mouse visual acuity works well when training encoders or decoders of vision on this dataset. Thus downsampling the video to the RL agent screen resolution of 86 x 155 meets ~0.4 deg/px = ~2.4x mouse visual acuity, and isn't expected to degrade encoder performance.}
Finally, validation/test split was accomplished by selecting one minute of samples from one session per mouse for the validation set, reserving the rest for testing. These validation periods were chosen during periods of high activity and sustained behavior performance at the full target distance. Note that these are provided as contiguous as possible, but each session's validation release will contain 1-2 time discontinuities that were unavoidable in order to remove perturbation periods. Each released data struct contains an additional \verb|time| vector with the exact session time in seconds per sample, so these discontinuities can be located if needed.

\bibliographystyle{plainnat}

\end{multicols}

\end{document}